\documentclass[aps,prd,showpacs, 
twocolumn] 
{revtex4}
\usepackage{graphicx} 
\usepackage{hyperref} 
\usepackage{amssymb, amsmath}
\usepackage{epsf}

\def\la{\; \raise0.3ex\hbox{$<$\kern-0.75em\raise-1.1ex\hbox{$\sim$}}\;}
\def\ga{\;  \raise0.3ex\hbox{$>$\kern-0.75em\raise-1.1ex\hbox{$\sim$}}\;}

\def\la{\; \raise0.3ex\hbox{$<$\kern-0.75em\raise-1.1ex\hbox{$\sim$}}\;}
\def\ga{\;  \raise0.3ex\hbox{$>$\kern-0.75em\raise-1.1ex\hbox{$\sim$}}\;}
\def\pFn{p_{\raise-0.3ex\hbox{{\scriptsize F$\!$\raise-0.03ex\hbox{\rm n}}}}
}  % p_Fn

\def\pFa{p_{\raise-0.3ex\hbox{{\scriptsize F$\!$\raise-0.03ex\hbox{$i$}}}}
}  % p_{F i}

\def\pFas{p_{\raise-0.3ex\hbox{{\scriptsize F$\!$\raise-0.03ex\hbox{$k$}}}}
}  % p_{F i}

\def\pFb{p_{\raise-0.3ex\hbox{{\scriptsize F$\!$\raise-0.03ex\hbox{$\beta$}}}}
}  % p_{F \beta}

\def\vFa{v_{\raise-0.3ex\hbox{{\scriptsize F$\!$\raise-0.03ex\hbox{$i$}}}}
}  % v_{F i}
\def\pFp{p_{\raise-0.3ex\hbox{{\scriptsize F$\!$\raise-0.03ex\hbox{\rm p}}}}
}  % p_Fp
\def\pFe{p_{\raise-0.3ex\hbox{{\scriptsize F$\!$\raise-0.03ex\hbox{\rm e}}}}
}  % p_Fe
\def\pFmu{p_{\raise-0.3ex\hbox{{\scriptsize F$\!$\raise-0.03ex\hbox{\rm
$\mu$}}}} }  % p_Fe
\def\m@th{\mathsurround=0pt }
\def\eqalign#1{\null\,\vcenter{\openup1\jot \m@th
   \ialign{\strut$\displaystyle{##}$&$\displaystyle{{}##}$\hfil
   \crcr#1\crcr}}\,}

\begin{document}

%%%%%%%%%%%%%%%%%%%%%%%%%%%%%%%%%%%%%%%%%%%%%%%%%%%%%%%%%%%%%%%%%%%%%%
\title{Decoupling of superfluid and normal modes in pulsating neutron stars}
%%%%%%%%%%%%%%%%%%%%%%%%%%%%%%%%%%%%%%%%%%%%%%%%%%%%%%%%%%%%%%%%%%%%%%
%
%
\author{Mikhail E. Gusakov$^{1}$ and Elena M. Kantor$^{1,2}$}
\affiliation{
$^1$Ioffe Physical Technical Institute,
Polytekhnicheskaya 26, 194021 St.-Petersburg, Russia
%e-mail: gusakov@astro.ioffe.ru
\\
$^2$St.-Petersburg State Polytechnical University,
Polytekhnicheskaya 29, 195251 St.-Petersburg, Russia
}
%\author{Mikhail E. Gusakov and Elena M. Kantor}
%%
%\affiliation{
%Ioffe Physical Technical Institute,
%Politekhnicheskaya 26, 194021 Saint-Petersburg, Russia}
%%e-mail: gusakov@astro.ioffe.ru}
%%\date{}
%%
%%\preprint{}

%%%%%%%%%%%%%%%%%%%%%%%%%%%%%%%%%%%%%%%%%%%%%%%%%%%%%%%%%%%%%%%%%%%%%%
\begin{abstract}
We show that equations governing 
pulsations of superfluid neutron stars (NSs)
can be split into two sets of weakly coupled equations,
one describing the superfluid modes and another one, the normal modes.
The coupling parameter $s$ is small, $|s| \sim 0.01-0.05$, 
for realistic equations of state (EOSs). 
Already an approximation $s=0$ is sufficient to calculate the 
pulsation spectrum within the accuracy of a few percent.
Our results indicate, in particular, 
that emission of gravitational waves 
from superfluid pulsation modes 
is suppressed in comparison to that from normal modes. 
The proposed approach allows to drastically simplify 
modeling of pulsations of superfluid NSs.
\end{abstract}
%%%%%%%%%%%%%%%%%%%%%%%%%%%%%%%%%%%%%%%%%%%%%%%%%%%%%%%%%%%%%%%%%%%%%%

\maketitle

{\it \bf Introduction}. ---
The pulsations of NSs
can be excited either 
due to internal instabilities 
or owing to external perturbations. 
Currently the detectors
that will be able, according to preliminary estimates, 
to register gravitational waves from pulsating NSs,
are under development \cite{and_etal09}.
For the correct interpretation of future observations
it is necessary to have a well developed theory of NS pulsations.
The formulation of such a theory is complicated
by the fact that at a temperature $T \la 10^8-10^{10}$~K,
baryons in the internal layers of NSs become superfluid.
Thus, to model pulsations 
one has to employ
superfluid hydrodynamics which is much more complicated 
than the ordinary one, 
describing ``normal'' (nonsuperfluid) matter.

For the first time the global pulsations of superfluid NSs
were analyzed by Lindblom and Mendell in 1994 \cite{lm94}.
Considering a simple model of a Newtonian star they numerically found 
two distinct classes of pulsation modes:
(i) normal modes which practically coincide with the
corresponding modes of a normal star; and,
(ii) superfluid modes in which the matter pulsates 
in such a way that the mass current density 
approximately vanishes.   
The subsequent numerical studies of various pulsation modes
(the literature is vast; see, e.g., Refs.\ \cite{puls, acl02} 
and references therein) 
confirmed the result of Ref.\ \cite{lm94} though
general explanation of this result 
has not yet been proposed \cite{snoska1}.    
In this work we give such an explanation.
In addition, we present an approximate scheme 
which allows to greatly simplify 
calculations of pulsating superfluid NSs.
In what follows the speed of light 
$c=1$.

{\it \bf Superfluid hydrodynamics}. ---
For simplicity, we consider NS cores composed of 
neutrons ($n$), protons ($p$), and electrons ($e$).
We also assume that protons are normal 
while neutrons are superfluid 
in some region of a NS core. 
As demonstrated in Ref.\ \cite{gk10} 
possible admixture of other particle species (e.g., muons)
and proton superfluidity do not affect our principal results.
Finally, we first consider a nonrotating NS.
Effects of rotation will be 
incorporated later in the text. 

It is well known that in superfluid matter, 
several independent motions 
with different velocities may coexist 
without dissipation \cite{khalatnikov89}.
In our case the system is fully described by 
two four-vectors, $u^{\mu}$ and $w_{(n)}^{\mu}$.
The vector $u^{\mu}$ is the velocity of
electrons and protons as well as ``normal'' neutrons; 
the vector $w_{(n)}^{\mu}$ arises from 
additional degrees of freedom associated 
with neutron superfluidity.
In the nonrelativistic limit 
the spatial components of the four-vector 
$w_{(n)}^{\mu}$
are related to superfluid velocities ${\pmb V}_{s n}$
of the Landau-Khalatnikov theory \cite{khalatnikov89} by the equality
${\pmb w_{(n)}}=m_n ({\pmb V}_{s n} - {\pmb u})$,
where $m_n$ is the neutron mass; 
${\pmb u}$ is the spatial component 
of the ``normal'' four-velocity $u^{\mu}$.
The electron $j_{(e)}^\mu$,
proton $j_{(p)}^\mu$, and neutron $j_{(n)}^\mu$ current densities
are expressed through the vectors $u^{\mu}$ and $w^{\mu}_{(n)}$
as \cite{ga06, gusakov07}: 
$j_{(e)}^{\mu}=n_e u^{\mu}$,
$j_{(p)}^{\mu}=n_p u^{\mu}$, and
$j_{(n)}^{\mu}=n_n u^{\mu} + Y_{nn} w^{\mu}_{(n)}$.
%
%
%\begin{equation}
%j_{(e)}^{\mu}=n_e u^{\mu}, \quad
%j_{(p)}^{\mu}=n_p u^{\mu}, \quad
%j_{(n)}^{\mu}=n_n u^{\mu} + Y_{nn} w^{\mu}_{(n)}.
%\label{currents}
%\end{equation}
%
Here $n_l$ is the number density 
of particles $l=n$, $p$, or $e$.
The expression for $j_{(n)}^{\mu}$ 
consists of two terms
reflecting the fact that both normal
and superfluid liquid components 
contribute to neutron current density.
The coefficient $Y_{nn}$ has been calculated 
in Ref.\ \cite{gkh09ab}; 
it is a relativistic analogue
of superfluid density of neutrons $\rho_{s n}$. 
%(see, e.g., Ref.\ \cite{khalatnikov89}).
In the nonrelativistic limit $Y_{nn}=\rho_{s n}/m_n^2$.
This coefficient depends on $T$ and increases steadily from 0
for normal matter (when $T \geq T_{cn}$, 
where $T_{cn}$ is the neutron critical temperature) 
to $n_n/\mu_n$ for entirely superfluid matter ($T=0$).
Here and below, $\mu_l$ is the chemical potential
for particles $l=n$, $p$, or $e$.

%In what follows 
To proceed further, we assume that:
(i) the quasineutrality condition holds both 
for equilibrium and pulsating matter, $n_p=n_e$;
and, (ii) an unperturbed NS is in beta-equilibrium, 
i.e. the disbalance of chemical potentials 
$\delta \mu \equiv \mu_n-\mu_p-\mu_e=0$.
It is convenient then to formulate 
the system of nondissipative 
hydrodynamic equations 
using the baryon number density $n_b \equiv n_n+n_p$  
and 
$\delta \mu$ 
as independent variables \cite{gusakov07}.
The system consists of 
(1) the continuity equations for baryons and electrons
\begin{equation}
j^{\mu}_{(b) \, ;\mu}=0, \quad \quad
j^{\mu}_{(e) \, ;\mu}=0,
\label{cont}
\end{equation}
where the baryon current density is
$j^{\mu}_{(b)} \equiv j^{\mu}_{(n)}+j^{\mu}_{(p)}=
n_b u^{\mu} + Y_{nn} w^{\mu}_{(n)}$;
(2) Einstein equations
\begin{equation}
R^{\mu \nu} - 
\frac{1}{2} 
%1/2
\, g^{\mu \nu} \, R 
=-8 \pi G\, \, T^{\mu \nu}
\label{Rmunu}
\end{equation}
with the energy-momentum tensor 
$T^{\mu \nu} = (P+\varepsilon) \, u^{\mu} u^{\nu} + P g^{\mu \nu}
+ Y_{nn} ( \underline{w^{\mu}_{(n)} w^{\nu}_{(n)}} + \mu_n \, w^{\mu}_{(n)} u^{\nu}$ 
$+ \mu_n \, w^{\nu}_{(n)} u^{\mu})$;
%
%\begin{equation}
%T^{\mu \nu} = (P+\varepsilon) \, u^{\mu} u^{\nu} + P g^{\mu \nu} 
%+ Y_{nn} \left( \underline{w^{\mu}_{(n)} w^{\nu}_{(n)}} + \mu_n \, w^{\mu}_{(n)} u^{\nu}
%+ \mu_n \, w^{\nu}_{(n)} u^{\mu} \right);
%\label{Tmunu}
%\end{equation}
%
(3) the potentiality condition for superfluid motion of neutrons
\begin{equation}
\partial_{\nu} \left[ w_{(n) \mu} + \mu_{n} u_{\mu} \right]
-\partial_\mu \left[ w_{(n) \nu} + \mu_{n} u_{\nu} \right]=0,
\label{potential}
\end{equation}
and (4) the second law of thermodynamics
\begin{equation}
d \varepsilon = \mu_n \, d n_b - \delta \mu \, d n_e + \underline{\underline{T \, dS}} 
+ \underline{Y_{nn} \, d \left( w^{\alpha}_{(n)} w_{(n) \alpha} \right)/2}. 
\label{2ndlaw}
\end{equation}
In Eqs.\ (\ref{Rmunu})--(\ref{2ndlaw}),
$R^{\mu \nu}$, $R$, and $g^{\mu \nu}$ are Ricci tensor, 
scalar curvature, and metric tensor, respectively;
$\partial_\mu \equiv \partial/\partial x^{\mu}$;
$G$ is the gravitation constant;
$P \equiv -\varepsilon + \mu_n \, n_b 
- \delta \mu \, n_e + \underline{\underline{T \, S}}$
is the pressure;
$\varepsilon$ and $S$ 
are the energy and entropy densities, respectively.
All the thermodynamic quantities are defined in the {\it comoving} frame 
in which $u^{\mu}=(1,0,0,0)$. 
This imposes an additional constraint on 
%the four-vector 
$w^{\mu}_{(n)}$ \cite{gusakov07}, 
$u_{\mu} \, w^{\mu}_{(n)}=0$. 
The solution to Eqs.\ (\ref{cont})--(\ref{2ndlaw})
in the superfluid region should be matched with that 
in the residual (normal) region of a star.
Equations, describing pulsations of normal matter can be
obtained from the system (\ref{cont})--(\ref{2ndlaw})
if one puts $Y_{nn}=0$ and ignores 
the condition (\ref{potential}).

%%%%%%%%%%%%%%%%%%%%%%%%%%%%%%%%%%%%%%%%%%%%%%%%%%%%
{\it \bf Linear Approximation}. ---
In this work we assume that in the unperturbed NS
${\pmb w}_{(n)}=0$, i.e. velocities 
of normal and superfluid components coincide.
For a nonrotating NS this simply means that both components are at rest.
%in the reference frame of a star.  
Then it follows
from the constraint $u_{\mu} \, w^{\mu}_{(n)}=0$
that 
%the four-vector 
$w^{\mu}_{(n)}$ vanishes in equilibrium, $w^\mu_{(n)}=0$ 
(because $u^{0} \neq 0$).
As a consequence, the terms in Eq.\ (\ref{2ndlaw}) 
and in the expression for $T^{\mu \nu}$,
underlined with one line, are quadratically small.
Similarly,
the terms which depend on $T$ and underlined twice,
are small in the strongly degenerate matter of NSs
and can be omitted \cite{gusakov07}.
%we also neglect them in the following.
Because of the same reasons one can consider, e.g.,   
the quantities $P$ and $\delta \mu$ 
as functions of only $n_b$ and $n_e$
(and neglect their dependence on scalars 
$w^{\mu}_{(n)} w_{(n) \mu}$ and $T$).

Now we make use of the energy-momentum conservation law 
$T^{\mu \nu}_{;\nu}=0$ which can be derived from Eq.\ (\ref{Rmunu}).
Composing a vanishing combination 
$T^{\mu \nu}_{;\nu}+u^{\mu} \, u_{\nu} \, T^{\alpha \nu}_{; \alpha}$
and subtracting from it Eq.\ (\ref{potential}) multiplied by $n_b \, u^{\mu}$,
one gets, with the help of Eq.\ (\ref{2ndlaw}) and the expression for $P$,
\begin{eqnarray}
&&-n_e \, \partial_\mu \delta \mu 
- u_{\mu} \, u^{\nu} \, n_e \, \partial_\nu \delta \mu
 -n_e \, \delta\mu \,\, u^{\nu} \, (u_{\mu})_{;\nu}
 \nonumber\\
 &&+ (g_{\mu \nu}+u_{\mu} u_{\nu}) \, u^{\alpha} \, X^{\nu}_{; \alpha} 
 + X^{\nu} \, u_{\mu \, ;\nu} + X_{\mu} \, u^{\nu}_{; \nu}
 \nonumber\\
 &&-n_b \, u^{\nu} \left[ \partial_\nu w_{(n) \mu} - \partial_\mu w_{(n) \nu}\right]=0.
% \nonumber\\
% &+& O(T^2) + O(w_{(n)}^{\mu} w_{(n)}^{\nu})=0.
\label{sfl}
\end{eqnarray}
Here $X^{\mu} \equiv \mu_n \, Y_{nn} \, w^{\mu}_{(n)}$.
The obtained ``superfluid'' equation is very attractive 
because
each term in 
it depends either 
on $\delta \mu$ or $w^{\mu}_{(n)}$.
Both these quantities are small in a slightly perturbed matter
(and vanish in equilibrium).
This means that 
in the linear approximation 
Eq.\ (\ref{sfl}) does not depend explicitly
on the perturbations of the metric $g_{\mu \nu}$ 
and 
the four-velocity $u^{\mu}$.
Thus, one can replace $g_{\mu \nu}$ and $u^{\mu}$ in Eq.\ (\ref{sfl})
by their equilibrium values.
For a nonrotating NS the spatial components of Eq.\ (\ref{sfl})
can be rewritten in a remarkably simple form ($j=1$, $2$, $3$)
\begin{equation}
i \, \omega \, 
(\mu_{n} \, Y_{nn} - n_{b}) \, w_{(n)j}
= n_{e} \,\, \partial_j (\sqrt{-g_{00}} \,\,\delta \mu),
\label{sfl1}
\end{equation}
where we assumed that 
$w^\mu_{(n)}$ depends on time $t$ as 
$w^\mu_{(n)} \sim {\rm exp}(i \omega t)$.
In Eq.\ (\ref{sfl1}) all the quantities except for 
$\delta \mu$ and $w^{\mu}_{(n)}$ are taken in equilibrium. 
Near the equilibrium, the function $\delta \mu(n_b, n_e)$
can be expanded in the Taylor series and presented, 
in the linear approximation, as 
\begin{equation}
\delta \mu = n_e \, (\partial \delta \mu/\partial n_e) \,
(z \, D_1+D_2).
\label{dmu}
\end{equation}
Here
$z \equiv [n_b \partial \delta \mu/\partial n_b]
/[n_e \partial \delta \mu/\partial n_e]$; $D_1 \equiv \delta n_b/n_b$; 
$D_2 \equiv \delta n_e/n_e$. 
The symbol $\delta$ in front of some quantity denotes 
a deviation of this quantity from its equilibrium value. 
The dimensionless functions $D_1$ and $D_2$
can be found from 
%the continuity equations 
Eq.\ (\ref{cont})
and depend on $w^{\mu}_{(n)}$, $u^{\mu}$, and $g_{\mu \nu}$.
Thus, generally, Eq.\ (\ref{sfl1})
is not independent and should be solved together with
Einstein equations (\ref{Rmunu}).
In the linear approximation Eq.\ (\ref{Rmunu})
%for a pulsating NS 
can be written in the following 
symbolic form: 
$\delta(R^{\mu \nu} - 1/2 \,\, g^{\mu \nu} \, R) 
=-8 \pi G\, \, \delta T^{\mu \nu}$. 
The left-hand side of this equation contains only 
perturbations of metric. 
To write out the right-hand side it is convenient 
to introduce new independent variables, 
the four-velocity of baryons
$U^{\mu} \equiv j^{\mu}_{(b)}/n_b=u^{\mu} 
+ Y_{nn} w^{\mu}_{(n)}/n_b$ 
and $W^{\mu} \equiv Y_{nn} w^{\mu}_{(n)}/n_b$,
instead of $u^{\mu}$ and $w^{\mu}_{(n)}$.
Note that for an unperturbed star $U^{\mu}=u^{\mu}$ and $W^{\mu}=0$
(since in equilibrium $w^{\mu}_{(n)}=0$).
The same is also true for a pulsating NS 
if the matter is nonsuperfluid (because then $Y_{nn}=0$).
Employing the new variables an expression for $\delta T^{\mu \nu}$
takes the form
\begin{eqnarray}
&&\delta T^{\mu \nu} = 
(\delta P + \delta \varepsilon) \, U^{\mu}  U^{\nu} 
\nonumber\\
&&+ 
(P + \varepsilon ) 
(U^{\mu} \, \delta U^{\nu}+ U^{\nu} 
\, \delta U^{\mu}) + \delta P \, g^{\mu \nu} 
+ P \, \delta g^{\mu \nu}. 
\label{deltaTmunu}
\end{eqnarray}
Here the quantities $P$, $\varepsilon$, $U^{\mu}$ and $g^{\mu \nu}$
are taken at equilibrium.
As follows from Eq.\ (\ref{2ndlaw}) the variation 
$\delta \varepsilon$ equals 
$\delta \varepsilon = \mu_n \delta n_b$
and depends on $\delta U^{\mu}$ and $\delta g^{\mu \nu}$
(and is independent of $\delta W^{\mu}$).
The variation $\delta P$ can be expanded in analogy with Eq.~(\ref{dmu}),
\begin{equation}
\delta P = n_b \, (\partial P/\partial n_b) \,
(D_1 + s \, D_2),
\label{dP}
\end{equation}
where the function $D_1$ depends on $\delta U^{\mu}$ and $\delta g^{\mu\nu}$,
%while 
and $D_2$ depends on the difference $(\delta U^{\mu}- \delta W^{\mu})$ 
and $\delta g^{\mu\nu}$ 
[see the definitions for $D_1$ and $D_2$, and Eq.\ (\ref{cont})].
The parameter $s$, hereafter referred to as
the ``coupling parameter'', is given by
$s \equiv (n_e \, \partial P/\partial n_e)
/(n_b \, \partial P/\partial n_b)$. 
%

%%%%%%%%%%%%%%%%%%%%%%%%%%%%%%%%%%%%%%%%%%%%%%%%%%%%%%%%%%%%%%%%%%
{\it \bf Superfluid and normal modes}. ---
If $s=0$ then 
Eq.\ (\ref{deltaTmunu}) for 
$\delta T^{\mu \nu}$
does not depend on $\delta W^{\mu}$ 
and has {\it exactly} the same form as in the absence of superfluidity.
In that case the Einstein equations (and boundary conditions to them)
coincide, in the linear approximation, with the corresponding equations 
for normal matter.
They can be solved separately from the ``superfluid'' Eq.\ (\ref{sfl1})
so that the solution 
(the spectrum of eigenfrequencies $\omega$ 
and the eigenfunctions $\delta U^{\mu}$) 
will be indistinguishable from that for a nonsuperfluid star.

%%%%%%%%%%%%%%%%%%%%%%%%%%%%%%%%%%%%%%%%%%%%%%%%%%%%%%%%%%%%%%%%%%%
\begin{figure}[t]
\setlength{\unitlength}{1mm}
\leavevmode
%\hskip  0mm
%\vskip  -3mm
%\includegraphics[width=47mm,bb=0 746 315 428,clip]{dPdne.eps}
%\includegraphics[width=47mm]{1.eps}
\includegraphics[width=75mm]{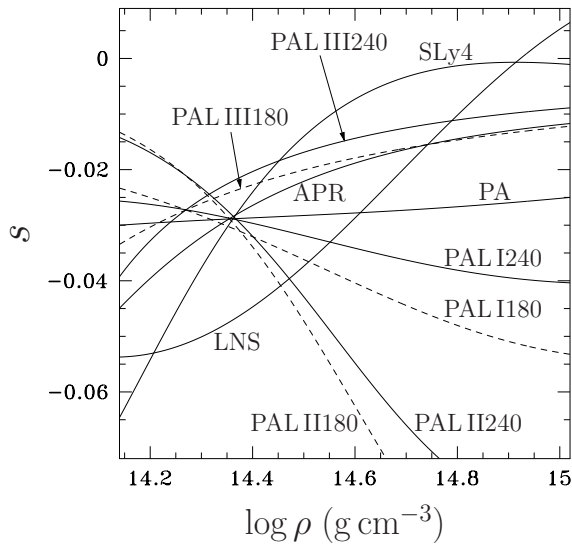}
%\vskip 0mm
\caption{
The coupling parameter $s$ versus density $\rho$ 
for EOSs PAL \cite{pal88}, PA \cite{pa92}, 
APR \cite{apr98}, SLy4 \cite{cbhms98}, and LNS \cite{clsg06}.
EOSs of the PAL family differ by the symmetry energy 
(models I, II, or III)
%; the notations are the same as in Ref.\ \cite{hly00})
and by the value of the compression modulus, $180$ or $240$ MeV.
The models with the compression modulus $120$ MeV 
are not plotted since they give 
too small maximum NS masses
that seem to contradict observations \cite{lp07}.
Note that the recently measured mass 
$M=(1.97 \pm 0.04) M_\odot$ \cite{dprrh10}
of the millisecond pulsar PSR J1614-2230
further rules out PA EOS and all PAL EOSs 
except for PAL I240 and PAL II240.
%Notice, that the recent mass measurement
%of the millisecond pulsar PSR J1614-2230
%gives $M=(1.97 \pm 0.04) M_\odot$ \cite{dprrh10} and 
%further rules out PA EOS and all PAL EOSs 
%except for PAL I240 and PAL II240.
}
\vskip -5mm
\label{fig1}
\end{figure}
%%%%%%%%%%%%%%%%%%%%%%%%%%%%%%%%%%%%%%%%%%%%%%%%%%%%%%%%%%%%%%%%%%%%
%
Let us now 
%discuss
focus on 
the following question.
Assume that $s$ still vanishes. 
Is it possible for a NS to oscillate on a frequency which
is not an eigenfrequency of a normal star?
Suppose that it is indeed the case.
Then the linearized Einstein equations will be satisfied 
only if $\delta U^{\mu}=0$ and $\delta g^{\mu \nu}=0$.
As follows from Eq.\ (\ref{cont}), in this case
$\delta n_b=0$ (i.e. $D_1=0$), 
while $D_2$ depends only on $\delta W^{\mu}$
(or, in other words, on $W^{\mu}$, since $W^{\mu}=0$ in equilibrium).
In particular, for a nonrotating NS
\begin{equation}
D_2= 
%\sum_{j=1}^3
[(\partial_j n_{e}/n_e) \, W^j  
+ W^{j}_{; j}]/(i \, \omega \, U^0).
\label{D2}
\end{equation}
Here $j=1$, $2$, and $3$; 
all the quantities, except for $W^j$,
are taken in equilibrium;
when calculating the covariant derivative 
one should use the unperturbed metric.
Eqs.\ (\ref{dmu}) and (\ref{D2}) allow to formulate Eq.\ (\ref{sfl1})
purely in terms of $W^j$.
A boundary condition to this equation, ${\pmb W}_{\bot}=0$,
also depends only on $W^{j}$ and can be obtained 
from the requirement that the baryon current density 
$j^{\mu}_{(b)}$ is continuous through the
normal-superfluid interface 
(${\pmb W}_{\bot}$ is the component of a vector $W^j$, 
perpendicular to the interface). 
Thus, Eq.\ (\ref{sfl1}) is self-contained and can be 
solved independently of Eq.\ (\ref{Rmunu}).
Its solution  
(eigenfrequencies and eigenfunctions $W^j$) 
describes superfluid modes
which were first considered in Ref.\ \cite{lm94} 
and do not have an analogue for a normal star.
To our best knowledge, 
the striking properties of such modes 
have not been discussed  
for a realistic model of a general 
relativistic NS 
at finite $T$.
First of all, the superfluid pulsation modes 
do not perturb metric ($\delta g^{\mu \nu}=0$)
and hence cannot emit gravitational waves.
In addition, because for these modes
$\delta U^{\mu}=0$ and $\delta n_b=0$,
the variations of $j_{(b)}^{\mu}$ and $P$ vanish, 
$\delta j_{(b)}^{\mu}=0$ and $\delta P=0$
[see Eqs.\ (\ref{cont}) and (\ref{dP})].
As a consequence, pulsations are localized entirely  
in the superfluid region of a star. 
%(and do not penetrate into the region where neutrons are normal).
In particular, they do not go to the NS surface.

%%%%%%%%%%%%%%%%%%%%%%%%%%%%%%%%%%%%%%%%%%%%%%%%%%%%%%%%%%%%%%%%%%%
\begin{figure*}[t]
\setlength{\unitlength}{1mm}
\leavevmode
%\hskip  0mm
%\includegraphics[width=136mm]{2.eps}
\includegraphics[width=175mm]{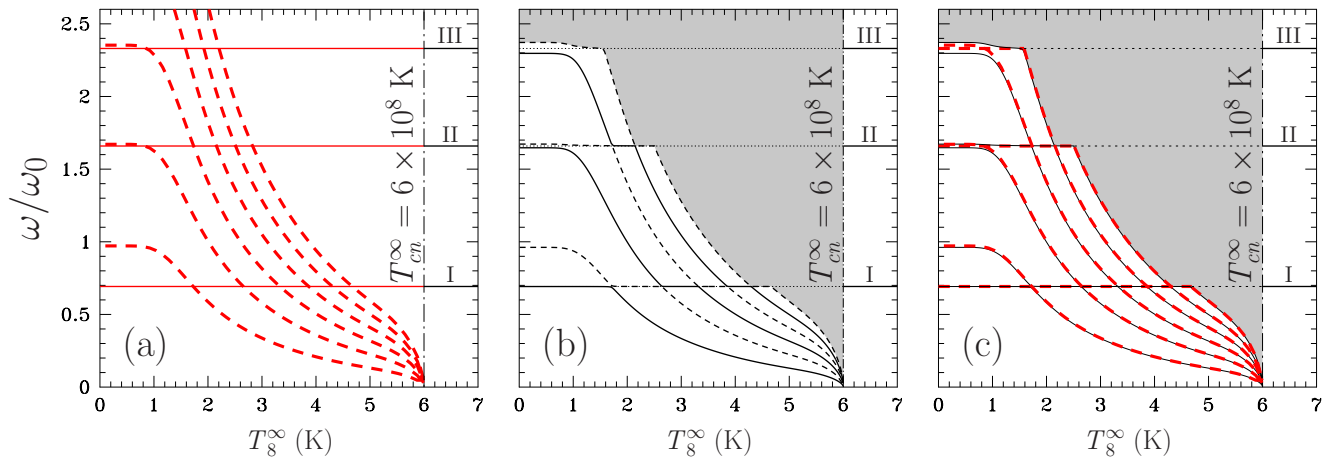}
\vskip -3mm
\caption{
Frequency $\omega$ in units of $\omega_0$ 
versus $T^\infty_8 \equiv T^{\infty}/(10^8$~K)
for various pulsation modes. 
(a) approximate spectrum; 
(b) exact spectrum;
(c) approximate (dashed lines) and exact (solid lines) spectra.
For more details see the text.
}
\vskip -5mm
\label{fig2}
\end{figure*}
%%%%%%%%%%%%%%%%%%%%%%%%%%%%%%%%%%%%%%%%%%%%%%%%%%%%%%%%%%%%%%%%%%%%
%
%
%%%%%%%%%%%%%%%%%%%%%%%%%%%%%%%%%%%%%%%%%%%%%
In the consideration above we supposed that $s=0$.
Yet, it is clear that  
superfluid and normal modes should remain 
approximately decoupled also at small but finite $s$. 
As follows from Fig.\ 1, 
$s$ is indeed small for realistic EOSs
and changes, on the average, from $-0.01$ to $-0.05$ \cite{snoska3}.
Taking into account that the parameter $z$ in Eq.\ (\ref{dmu})
is $z \sim -1$ 
for the same EOSs,
it is easy to show that for normal modes $D_1 \ga D_2$ 
[then the second term in Eq.\ (\ref{dP}) 
is much smaller than the first one],
while for superfluid modes $D_1 \sim s \, D_2$
[then the first term in Eq.\ (\ref{dmu}) 
is much smaller than the second one]. 
Generally, the exact solution 
of linearized pulsation equations (\ref{Rmunu}) and (\ref{sfl1}) 
can be presented 
as a series 
in 
%small 
parameter $s$ \cite{gk10}.
However, 
since $s$ is very small, 
the approximation 
of noninteracting Eqs.\ (\ref{Rmunu}) and (\ref{sfl1})
considered above 
(hereafter ``zero approximation'')   
is already sufficient to calculate the pulsation spectrum 
within the accuracy $\sim s$ (i.e., a few percent).

{\it \bf Example: Radial pulsations}. ---
Let us illustrate the obtained results with an example of 
a radially pulsating NS with the mass $M=1.4 M_{\odot}$.
We consider a simple NS model which was analyzed in detail
in Ref.\ \cite{ga06}.
In that paper it was assumed that the redshifted critical temperature 
of neutrons $T_{cn}^\infty$ is constant throughout the stellar core,
$T_{cn}^\infty=6 \times 10^8$ K. 
The results of approximate calculation of pulsation spectrum
are illustrated in Fig.\ 2a. 
The spectrum is calculated in zero approximation in $s$. 
%\cite{snoska2}.
In the figure, the pulsation frequency $\omega$
(in units of $\omega_0 \equiv c/R_{\rm NS}$, where 
$R_{\rm NS}=12.17$ km is the circumferential radius of a star)
is plotted as a function of internal redshifted stellar temperature 
$T^{\infty}$ for 3 normal (solid lines) and 6 superfluid (dashes)
pulsation modes. 
At $T^\infty > T_{cn}^\infty$ 
only the normal modes (I, II, and III) survive since
then the star is nonsuperfluid.
For comparison, in Fig.\ 2b we present the exact solution 
to the system of linearized equations (\ref{Rmunu}) and (\ref{sfl1}).
The first 6 modes are shown 
by alternate solid and dashed lines.
The spectrum was not plotted in the shaded region.
All other notations are the same as in Fig.\ 2a.
It is easy to see that the structure of both spectra is very similar.
However, there is one principal difference.
Instead of crossings of superfluid and normal modes in Fig.\ 2a, 
we have {\it avoided crossings} in Fig.\ 2b.
At these points the superfluid mode becomes normal and vice versa.
Such avoided crossings are not described in approximate treatment (Fig.\ 2a)
because when frequencies of superfluid and normal modes are close 
to each other,
Eqs.\ (\ref{Rmunu}) and (\ref{sfl1}) become strongly interacting
and cannot be considered as independent.
For comparison, we plot both spectra in Fig.\ 2c.
The exact solution is shown by solid lines, 
dashes correspond to the approximate solution. 
Other notations are the same as in Figs.\ 2a and 2b.
On average, the approximate solution differs from the exact one 
by $\sim 1.5-2 \%$.
For normal modes the difference becomes smaller with increasing of $T$.
In this case the number of ``superfluid'' neutrons decreases ($Y_{nn} \rightarrow 0$),
consequently, $W^j \equiv Y_{nn} \, w^j_{(n)}/n_b \rightarrow 0$ 
and zero approximation 
%\cite{snoska2} 
works better and better.

{\it \bf Taking into account rotation}. ---
Rotation leads to formation of Feynman-Onsager vortices 
inside NSs with the interspacing distance 
$\sim 10^{-2}-10^{-4}$ cm. 
The hydrodynamic equations averaged over the volume 
containing large amount of vortices
formally have the same form 
as in their absence \cite{bk60ml91} 
(if we neglect the small contribution 
of vortices to the internal energy density of matter). 
The only exception is the potentiality condition (\ref{potential})
that should be replaced by
$u^{\nu} \, \{\partial_{\mu} [  w_{(n) \, \nu} + \mu_n \, u_{\nu}]$
$- \partial_{\nu} [  w_{(n) \, \mu} + \mu_n \, u_{\mu}]
\}=O_{\mu \nu} \, W^{\nu}$,
where the tensor $O_{\mu \nu}$ 
is specified in Ref.\ \cite{gk10} and
is responsible for the interaction
between the normal and superfluid component. 
It can be found from the requirement that the entropy 
does not decrease.
Because of this 
%new 
condition the new term 
$n_b \,\, O_{\mu \nu} \, W^{\nu}$ appears
in the right-hand side of Eq.\ (\ref{sfl}).
Since this term depends on a small quantity $W^{\mu}$,
all our reasoning about decoupling of superfluid and normal modes
remain valid for rotating NSs as well. 

%%%%%%%%%%%%%%%%%%%%%%%%%%%%%%%%%%%%%%%%%%%%%%%%%%%%%%%%%%%%%%%%%%%%%%%%%%%%%
{\it \bf Comparison with previous works}. ---
%%%%%%%%%%%%%%%%%%%%%%%%%%%%%%%%%%%%%%%%%%%%%%%%%%%%%%%%%%%%%%%%%%%%%%%%%%%%%
For comparison we choose two papers, Refs.\ \cite{pr02} and \cite{acl02},
since at first sight it is not clear 
whether our results complement or contradict 
the conclusions drawn in these references.

The authors of Ref.\ \cite{pr02} 
considered a model of Newtonian star at $T=0$.
They demonstrated that 
superfluid modes decouple from the normal modes only
for an idealized case of nonstratified NSs, 
for which $n_e/n_b={\rm const}$ throughout the stellar core.

This result does not contradict ours because one can show
that the neutron-star matter is nonstratified 
only if $\partial P(n_b, n_e)/\partial n_e=0$ (that is $s=0$).
As follows from our analysis, 
in the latter case superfluid and normal modes 
are indeed strictly decoupled. 

The second conclusion made in Ref.\ \cite{pr02} 
is based on the observation 
that for most of the neutron-star models 
the stellar matter is stratified.
%Using for illustration their EOS II,
Using this observation
the authors of Ref.\ \cite{pr02} argued
that generally there should be no clear distinction between 
the superfluid and normal modes, 
or, in other words,
% that is
equations describing superfluid- and normal-type 
pulsations are strongly interacting.

This conclusion is not correct because,
as we demonstrated earlier in this work, 
the real coupling parameter $s$
can be small even for strongly stratified NSs 
(and is indeed small for realistic EOSs).

Now let us discuss the results of Ref.\ \cite{acl02}. 
This paper analyzed gravitational radiation
from superfluid nonrotating NSs at $T=0$
in the frame of the general relativity.
It was argued that superfluid modes 
must radiate gravitational waves in practically all situations, 
with intensity of radiation comparable to that from the normal modes
(unless an EOS has a very specific form 
satisfying Eq.\ (74) of Ref.\ \cite{acl02}).

When modeling the neutron-star pulsations 
the authors of Ref.\ \cite{acl02} used toy-model
EOSs that give completely 
unrealistic values for the coupling parameter $s$.
In particular, we found that their most 
realistic model II gives 
$s \sim 0.1$ at the center 
and $s= \infty$ at the superfluid-normal interface. 
Moreover, because their EOSs are artificial,
they were forced to relax an assumption of chemical equilibrium in the core.
As it is demonstrated in the present paper, 
the latter assumption is very important for the decoupling of modes 
and cannot be ignored.
Thus, it is not surprising that 
our results disagree with the results of Ref.\ \cite{acl02};
when $s$ is not small, 
superfluid modes can be as effective in radiating 
gravitational waves as normal modes.

In the end, 
it is worth mentioning
one more result of Ref.\ \cite{acl02}.
In that paper it is claimed
that any (nonradial)
pulsation mode must emit gravitational waves unless an EOS satisfies
some specific criterion [their Eq. (74)].
We checked that this criterion is not equivalent and does not follow
from our criterion $s=0$, which is a necessary condition 
for decoupling of superfluid modes from metric. 

{\it \bf Conclusion}. ---
Summarizing, equations describing pulsations of superfluid NSs
can be split into two systems of weakly coupled equations.
The coupling parameter $s$ 
of these systems is small for realistic EOSs, 
$|s| \sim 0.01-0.05$.
One system of equations describes normal modes, 
another one -- superfluid modes.
Already zero approximation in parameter $s$ 
%\cite{snoska2}
(when the systems are fully decoupled)
is sufficient to calculate the pulsation spectrum 
with an accuracy of a few percent.
In this approximation the normal modes coincide
with ordinary modes of nonsuperfluid NS, 
while superfluid modes do not perturb metric, pressure, 
baryon current density 
and are localized in superfluid region of a star.
%Let us notice, 
Note that an emission of gravitational waves
by superfluid modes is possible only 
in the next (first) order of perturbation theory in $s$.
Thus, it should be suppressed in comparison to gravitational 
radiation from the normal modes. 

Our finding that 
superfluid modes do not appear at the NS surface 
and do not emit gravitational waves in the $s=0$ limit
indicate that these modes should be 
very difficult to observe at small but finite $s$.
This means that observational properties 
of a pulsating superfluid star and a normal star of the same mass
should be very similar, so that it will be 
very hard to discriminate one from the other.

The obtained results explain numerical calculations \cite{lm94, puls}
and suggest simple perturbative (in parameter $s$) scheme
which drastically simplifies the problem of calculation 
of the pulsation spectrum for superfluid NSs. 
The presented approach allows to easily take into account
realistic EOSs, dissipation, 
various composition of matter, 
temperature effects, 
baryon superfluidity, density-dependent profiles of critical temperatures, 
and rotation of NSs. 
In more detail these issues will be discussed 
elsewhere \cite{gk10}.

{\it \bf Acknowledgements}. --- 
We thank D.P. Barsukov, A.I. Chugunov, 
and D.G. Yakovlev 
for valuable comments.
This research was supported by the Dynasty Foundation, 
Ministry of Education and Science of Russian Federation 
(Contract No. 11.G34.31.0001 
with SPbSPU and leading scientist G.G. Pavlov), 
RFBR (Grant No. 11-02-00253-a), 
and by FASI (Grant No. NSh-3769.2010.2).

\end{document}